\documentclass[twocolumn,prb]{revtex4}
\usepackage{epsfig,amssymb,amsmath}
\begin{document}
\title{Influence of coupling between junctions on breakpoint current in intrinsic Josephson junctions}

\author{ Yu.M.Shukrinov~$^{1,2}$}
\author{F.Mahfouzi~$^{2}$ }

\address{$^{1}$ BLTP, JINR, Dubna, Moscow Region, 141980, Russia \\
$^{2}$Institute for Advanced Studies in Basic Sciences, P.O.Box 45195-1159, Zanjan, Iran}

\begin{abstract}
We study theoretically the current voltage characteristics of intrinsic Josephson junctions in high-$T_c$
superconductors. An oscillation of the breakpoint current on the outermost branch  as a function of  coupling
$\alpha$ and dissipation $\beta$ parameters is found. We explain this oscillation as a result of the creation of
longitudinal plasma waves at the breakpoint with different wave numbers. We demonstrate the commensurability
effect and predict a group behavior of the current-voltage characteristics for the stacks with a different
number of junctions. A method to determine the wave number of longitudinal plasma waves from $\alpha$- and
$\beta$-dependence of the breakpoint current is suggested. We model the $\alpha$- and $\beta$-dependence of the
breakpoint current and obtain good agreement with the results of simulation.
\end{abstract}

\maketitle Creating new materials with given properties is an actual problem of physics, chemistry, and material
science. This is related to the system of Josephson junctions, too, which is a perspective object for
superconducting electronics and is being investigated intensively  now. A simulation of the current-voltage
characteristics (IVC) of a stacks of intrinsic Josephson junctions (IJJ)\cite{muller} at different values of the
model parameters such as the coupling and dissipation parameters is a way to predict the properties of the IJJ.
McCumber and Steward have investigated the return current as a function of dissipation parameter in a single
Josephson junction a long time ago.\cite{schmidt} In the case of the system of junctions, the situation is
cardinally different. The IVC of IJJ is characterized by a multiple branch structure and  branches have a
breakpoint region with its breakpoint current (BPC) and transition current to another branch.
\cite{sm-sust1,prb} The BPC is determined by the creation of the longitudinal plasma waves (LPW) with a definite
wave number $k$, which depends on the parameters $\alpha$ and $\beta$, the number of junctions in the stack, and
boundary conditions. If we neglect the coupling between junctions, the branch structure disappears, and the BPC
coincides with the return current. As we know, an investigation of the McCumber-Steward dependence for the
different branches of IVC for IJJ has not been done yet. Machida and Koyama\cite{machida04} have stressed that
capacitive coupling takes various values in HTSC and layered organic superconductors and they presented a
systematic study for the capacitively coupled Josephson junctions (CCJJ) model, focusing on the dependence of
phase dynamics on the strength of the capacitive coupling constant from weak to strong coupling regimes. But
they did not investgate the breakpoint region in the simulated IVC.

In this Letter, we generalized the McCumber-Steward dependence of the return current for the case of IJJ in the
HTSC. We  investigate the BPC $I_{bp}$ on the outermost branch  as a function  of the coupling $\alpha$ and
dissipation $\beta$ parameters for the stacks with a different number of IJJ and demonstrate a plateau with BPC
oscillation. Based on the idea of the parametric resonance in the stack of IJJ, a modeling of the
$\alpha\beta$-dependence of the BPC has been done, and good qualitative agreement with the results of simulation
has been obtained. We show that the $\alpha\beta$-dependence of the BPC is an instrument to determine the mode
of LPW created at the breakpoint in the stacks with a different number of junctions.

A system of dynamical equations in the capacitively coupled Josephson junctions model with diffusion current
(CCJJ+DC model)\cite{machida00,sm-physC2}
\begin{eqnarray}
\frac{d^2}{dt^2}\varphi_{l}=(I-\sin \varphi_{l} -\beta\frac{d\varphi_{l}}{dt})+ \alpha
(\sin \varphi_{l+1}+ \sin\varphi_{l-1} \nonumber \\- 2\sin\varphi_{l})+ \alpha
\beta(\frac{d\varphi_{l+1}}{dt}+\frac{d\varphi_{l-1}}{dt}-2\frac{d\varphi_{l}}{dt})
\label{d-phi-dif}
\end{eqnarray}
for the gauge-invariant phase differences $\varphi_l(t)=
\theta_{l+1}(t)-\theta_{l}(t)-\frac{2e}{\hbar}\int^{l+1}_{l}dz A_{z}(z,t)$  between superconducting layers
($S$-layers) for the stacks with a different number of intrinsic junctions has been numerically solved. Here
$\theta_{l}$ is the phase of the order parameter in S-layer $l$, $A_z$ is the vector potential in the barrier.

The CCJJ+DC model is different from the CCJJ model\cite{koyama96,matsumoto99,sm-physC1} by the last term on the
right hand side. This coupled Ohmic dissipation term might be derived by the microscopic theory\cite{machida00}
or phenomenologically by the inclusion of the diffusion current between S-layers and leads to the equidistant
branch structure in the IVC.\cite{sm-physC2} The details concerning the system ~(\ref{d-phi-dif}) are presented
in Ref.\cite{sm-physC2} Here we use the periodic boundary conditions considering the first S-layer as a neighbor
of the last one.

\begin{figure}[!ht]
 \centering
\includegraphics[height=40mm]{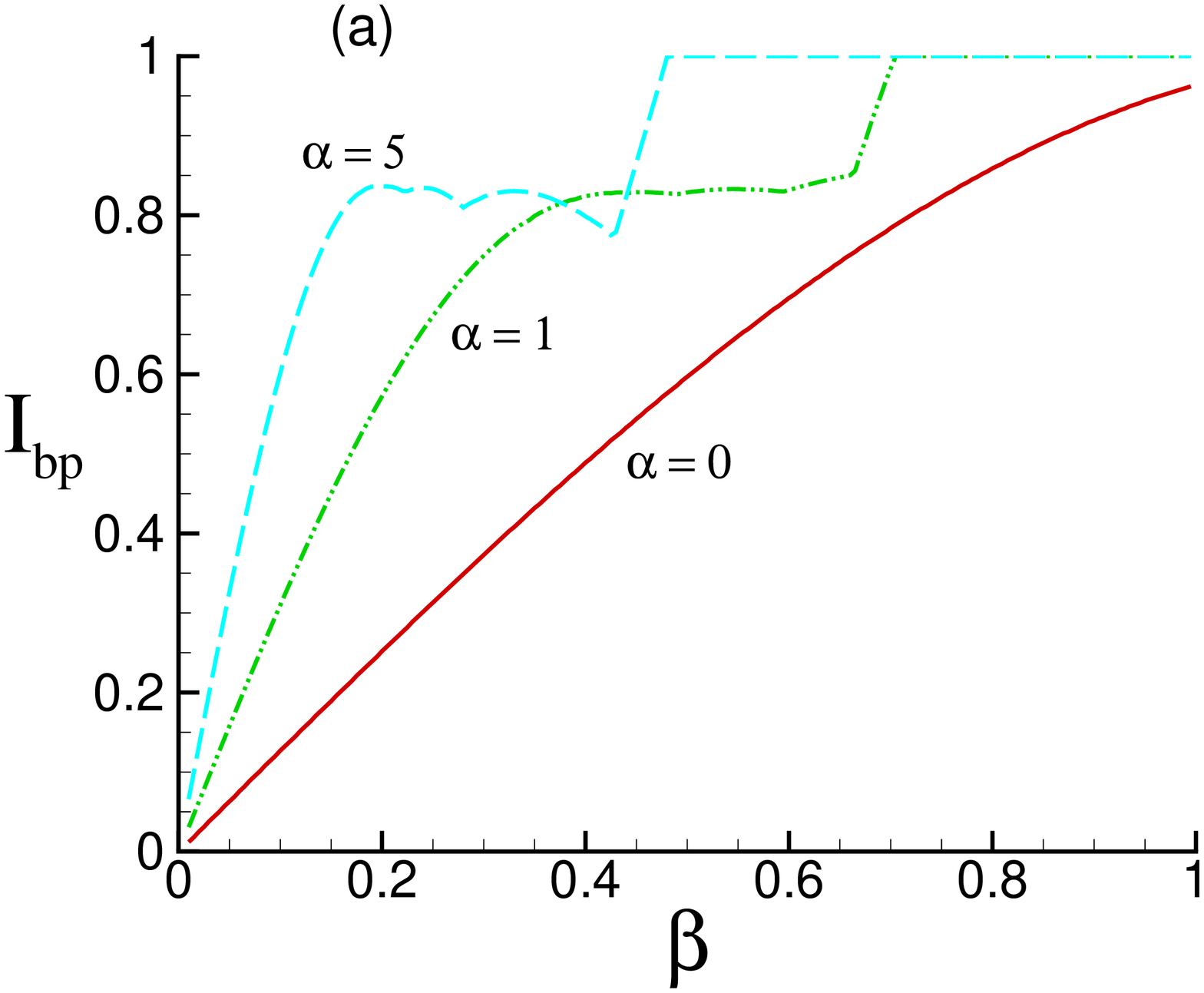}
\includegraphics[height=40mm]{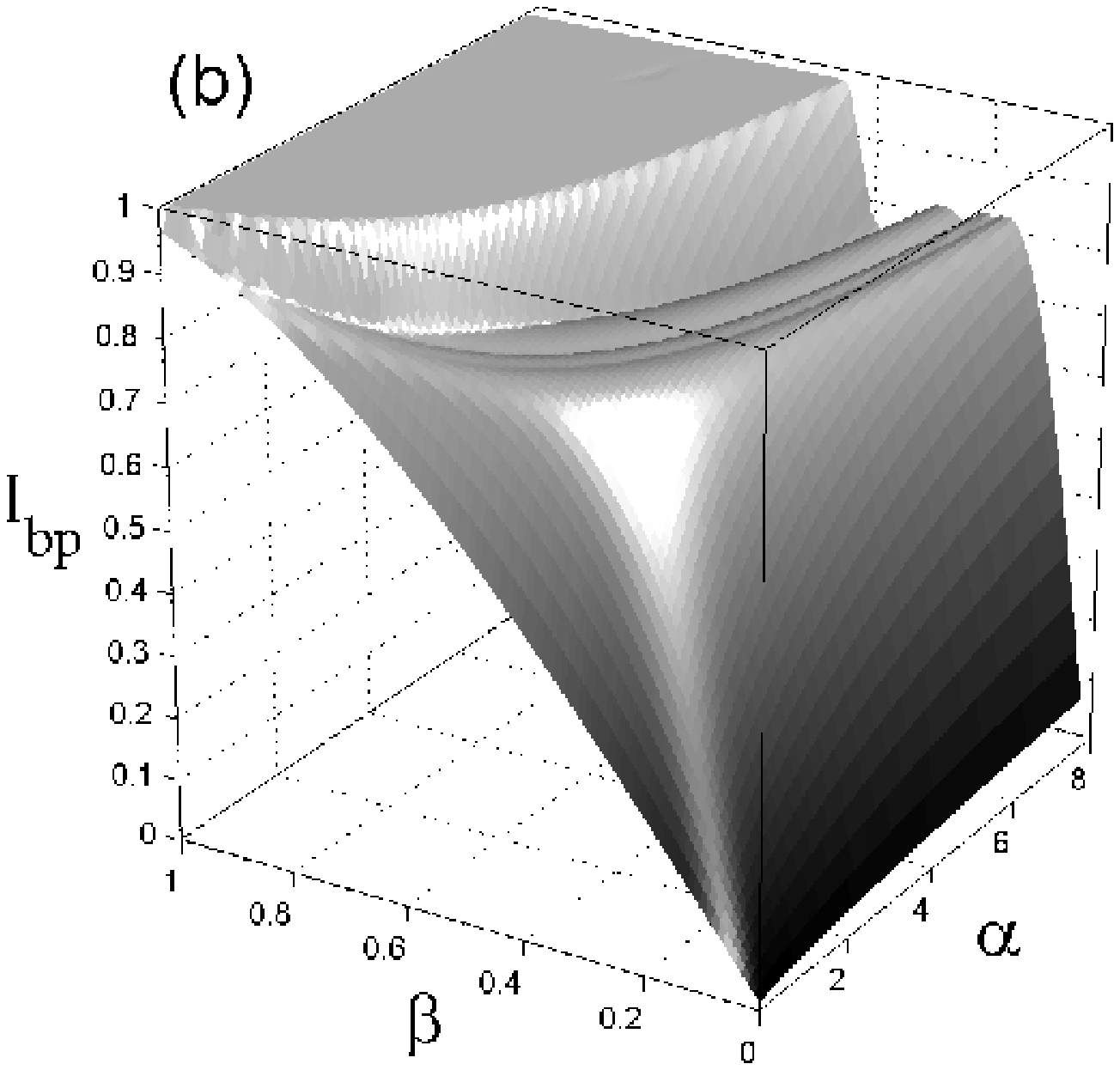}
\caption{(Color online) (a) - The $\beta$-dependence of the BPC $I_{bp}$ of the outermost branch in the IVC at
different values of coupling parameter $\alpha$; b) - The $\alpha \beta$-dependence of the $I_{bp}$  for a stack
of 10 IJJ.}
 \label{1}
\end{figure}

The simulated IVC have the breakpoint on their outermost branches. We have calculated the $\beta$-dependence of
the BPC $I_{bp}$ at fixed value of $\alpha$, changing $\beta$ in the interval (0,1) by step 0.005. The result of
the calculation at $\alpha = 0, 1 $ and 5 is presented in Fig.~\ref{1}a.
At $\alpha = 0$, the IVC does not manifest the multibranch structure, and the breakpoint coincides with the
return current.  The curves at $\alpha \neq 0$ have  new features in comparison with the case without coupling.
Particularly, they show a stronger increase of the $I_{bp}$  at small $\beta$, a plateau at $I_{bp} \simeq 0.83$
and the oscillation of the $I_{bp}$ on this plateau, and a transition to the non-hysteretic regime ( second
plateau) at smaller $\beta$. These features are discussed below. We change the coupling parameter $\alpha$ in
the interval (0,8) by step 0.1 and repeat the calculations of the $\beta$-dependence of $I_{bp}$. By this
method, we build the three-dimensional picture of the $\alpha\beta$-dependence of the $I_{bp}$  for a stack with
10 IJJ, which is shown in Fig.~\ref{1}b. We see two plateaus on this dependence and the oscillations of the
$I_{bp}$ on the first one as a function of $\alpha$ and $\beta$.
 We note the next features for the $\beta$-dependence : i) At $\alpha$ equal to
zero, our results for $\beta$-dependence of the $I_{bp}$ coincide with the previous simulation of the
$\beta$-dependence of the return current\cite{schmidt}; ii) at small $\beta$, the $\beta$-dependence is getting
sharper with the increase in $\alpha$; iii) the oscillations of the $I_{bp}$ are getting stronger at larger
$\alpha$; iiii) with the increase in $\alpha$, the transition to the non-hysteretic regime (to the second
plateau) is approached at smaller $\beta$ . For the  $\alpha$-dependence of the $I_{bp}$ we may note: i) At
small $\beta$, the $\alpha$-dependence is monotonic, and $I_{bp}$ is increasing with $\alpha$; ii) at some
$\beta$, the oscillations of $I_{bp}$ appear, iii) with the increase in $\beta$, the transition to the
non-hysteretic regime is observed  at smaller $\alpha$. The value of the $I_{bp}$ changes strongly at small
$\alpha$ and $\beta$. On the first plateau, the variation of the $I_{bp}$ consists of $\simeq 3\div 4 $ percent
of the value of $I_{c}$ for $N=10$. As we can see below, it depends on the number of junctions in the stack and
decreases with N.
\begin{figure}[!ht]
 \centering
\includegraphics[height=60mm]{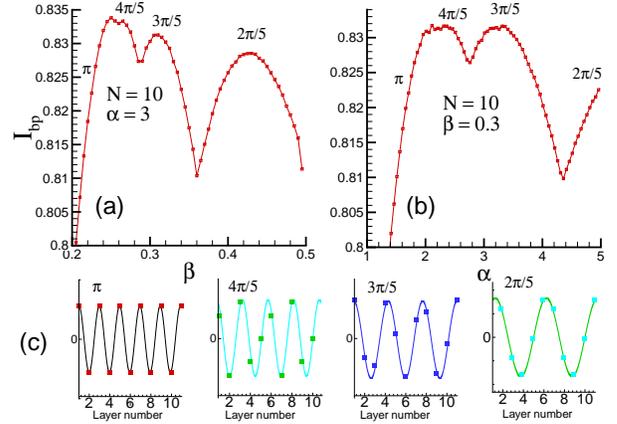}
\caption{(Color online) a) - The $\beta$-dependence of the $I_{bp}$ for a stack with 10 IJJ at $\alpha = 3$; b)
- The $\alpha$-dependence of the $I_{bp}$ at $\beta = 0.3$; c) - The charge distribution among the layers
corresponding to the  different plasma modes  in the stack of 10 IJJ at $\alpha = 3$  and $\beta = 0.24, 0.27,
0.3, 0.4$.}
 \label{2}
\end{figure}

Let us analyze in more detail the $\alpha$- and $\beta$-dependence of the $I_{bp}$. Fig.~\ref{1}a demonstrates
the general features of $\beta$-dependence of the $I_{bp}$ at different values of the coupling parameter. To
clearly show these features,  we demonstrate in Fig.~\ref{2}a in an increased scale the $\beta$-dependence of
the $I_{bp}$ at $\alpha = 3$. We can see clearly four maximums of $I_{bp}$ on this curve. Using the Maxwell
equation $div (E/d) =4\pi \rho$, we express the charge $\rho_i$  on the superconducting layer $i$ by the
voltages $V_{i,i-1}$ and $V_{i,i+1}$ in the neighbor insulating layers $\rho_i=\frac{\epsilon_0}{4\pi
d_0d}(V_{i, i+1}-V_{i-1 ,i})$. Solution of the system of equations ~(\ref{d-phi-dif}) gives us the voltages
$V_{i,i+1}$ in all junctions in the stack, and it allows us to investigate the time dependence of the charge on
each S-layer. We analyze the time dependence of the charge oscillations on S-layers at $\beta$ equal to 0.24,
0.27, 0.3 and 0.4 (around each maxima). The charge distributions among the S-layers in the stack at a fixed time
moment at the breakpoint of the outermost branch are presented in Fig.~\ref{2}c. The charge oscillations on
S-layers correspond to standing LPW with $k$ equal to $\pi$, $4\pi/5$, $3\pi/5$ and $2\pi/5$, relating to the
four different intervals of the $\beta$ with four maximums in this region. Fig.~\ref{2}b shows the
$\alpha$-dependence of $I_{bp}$ at $\beta = 0.3$, and it demonstrates four regions corresponding to the
different modes of LPW.

To prove our results and test the idea that at the breakpoint a parametric resonance is approached and plasma
mode is excited by  Josephson  oscillations, we have modeled the $\alpha\beta$-dependence of the $I_{bp}$ in the
CCJJ+DC model. The equation for the Fourier component of the difference of phase differences $\delta \varphi_{l}
= \varphi_{l+1,l}-\varphi_{l,l-1} $ between neighbor junctions is\cite{sm-sust1}
$\ddot{\delta_k}+\beta(k)\dot{\delta_k}+ \cos(\Omega(k)\tau)\delta_k=0$, where $\tau=\omega_p(k)t$,
$\omega_p(k)=\omega_p C$, $\beta(k)=\beta C$, $\Omega(k)=\Omega/C$ and $C = \sqrt{1+2\alpha(1-\cos(k))}$. This
equation shows a resonance with changing its parameters $\beta(k)$ and $\Omega(k)$.
\begin{figure}
 \centering
\includegraphics[height=35mm]{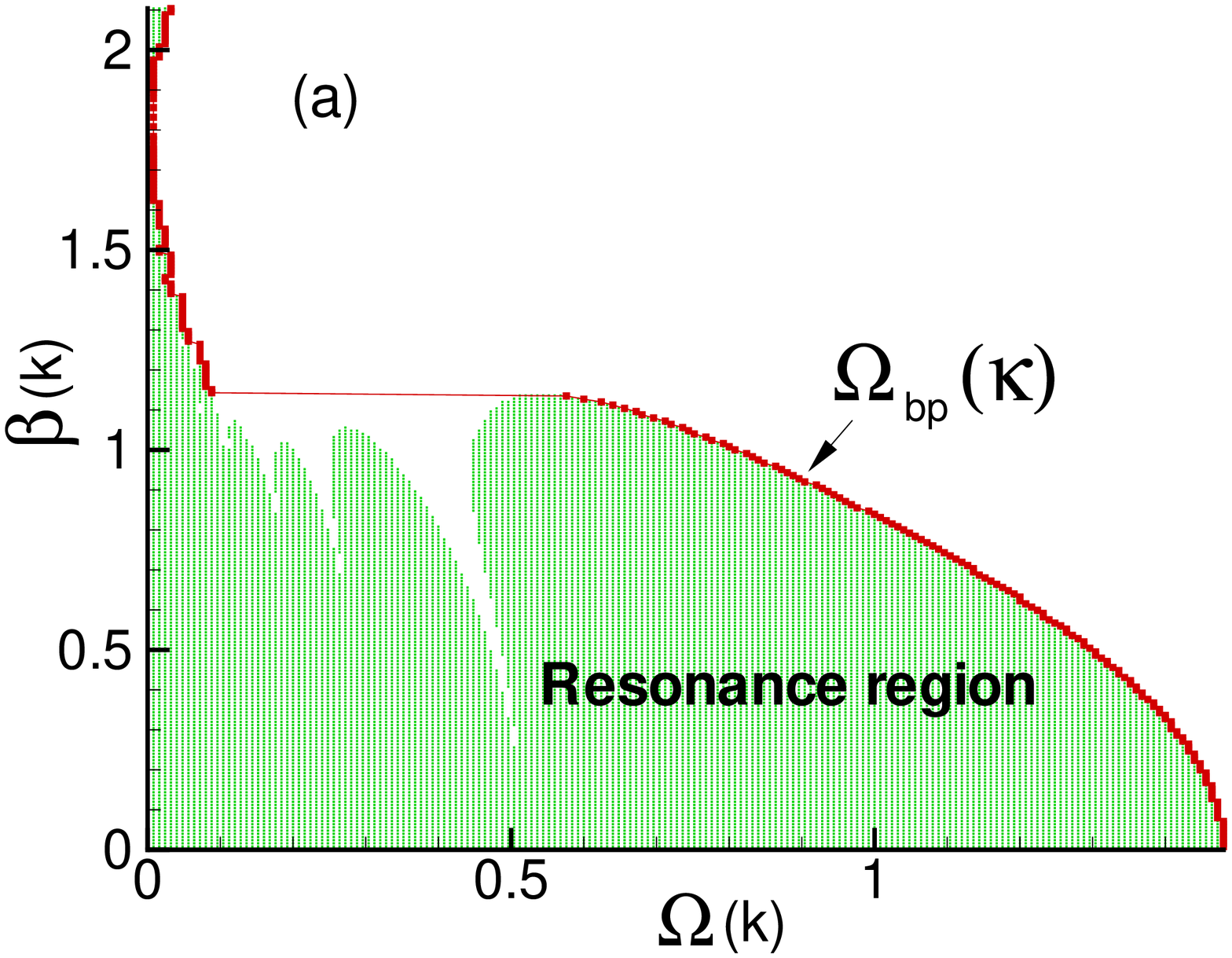}\includegraphics[height=35mm]{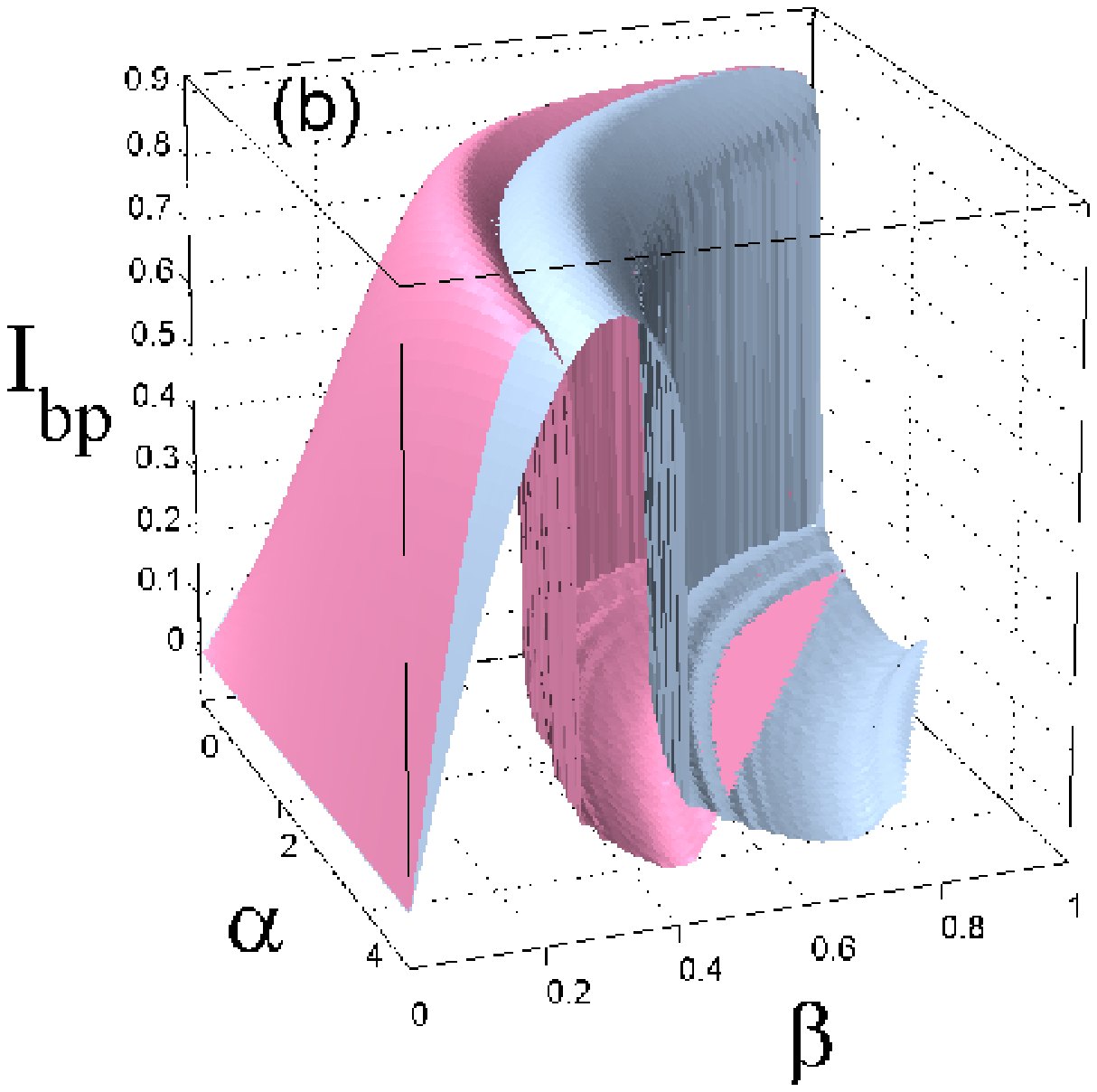}
\includegraphics[height=35mm]{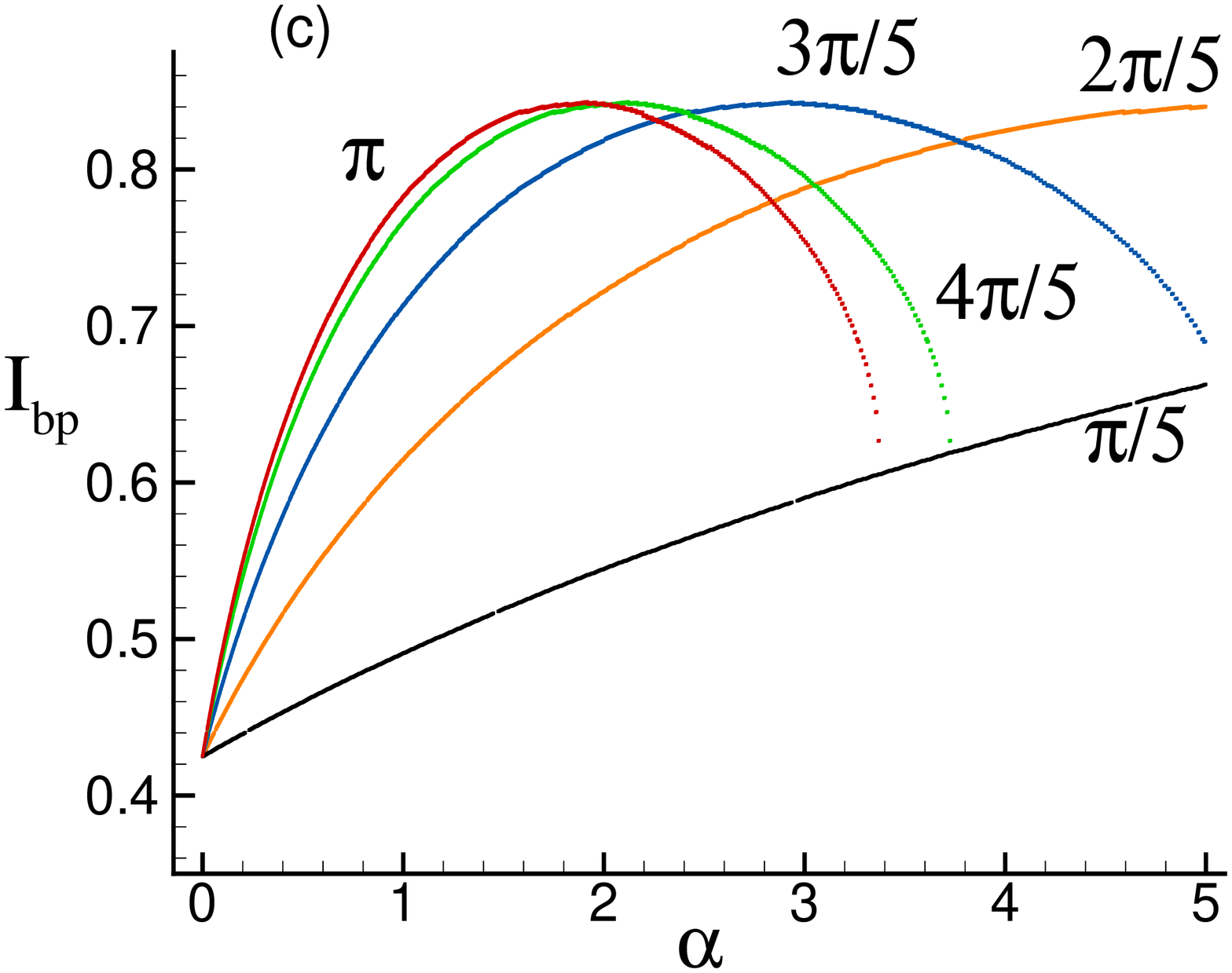}\includegraphics[height=35mm]{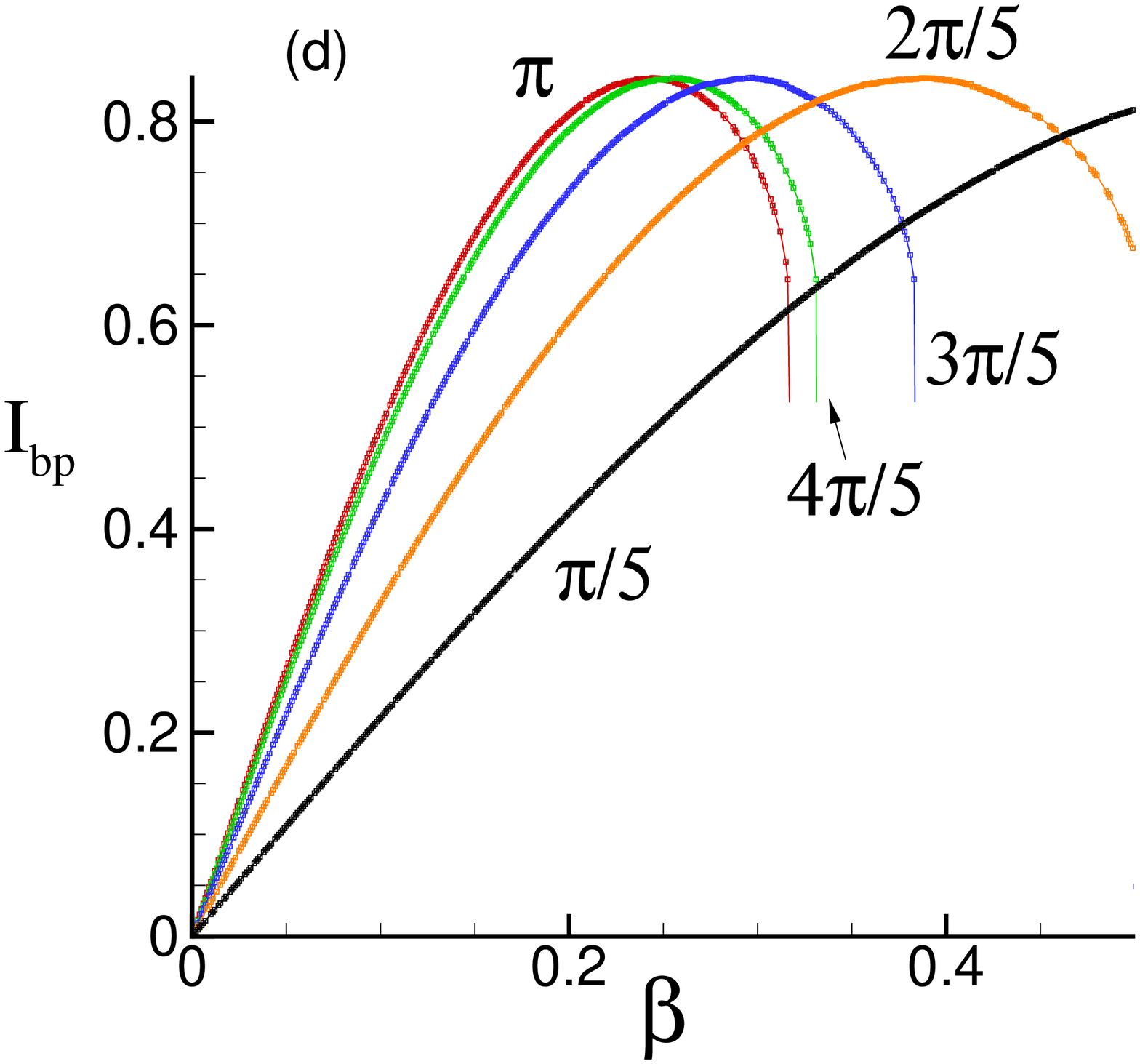}
\caption{ (Color online)  a) - Parametric resonance region  in $\Omega(k)-\beta(k)$ diagram. The value
$\Omega(k) = \Omega _{bp}(k)$ corresponds to the breakpoint voltage on the outermost branch; b) - Result of
modeling of the $\alpha\beta$-dependence of the $I_{bp}$ for plasma modes with $k=\pi$ and $k=2\pi/5$ for a
stack of 10 IJJ; c) - The modeled $\alpha$-dependence of $I_{bp}$ for stack with 10 IJJ at $\beta = 0.3$
corresponding to the creation of the LPW with different $k$; d) - The modeled $\beta$-dependence of $I_{bp}$ at
$\alpha = 3$.}
 \label{3}
\end{figure}
In Fig.~\ref{3}a, we have plotted the parametric resonance region for this equation  on the diagram
$\beta(k)-\Omega(k)$. Using this diagram, we determine the curve which corresponds to the edge of the resonance
region.  This curve is shown in Fig.~\ref{3}a by dots. We consider that the point on this curve corresponding to
$max \Omega (k)$ at a fixed value of $\beta (k)$ gives us the value of the $\Omega_{bp} (k)$ which corresponds
to the breakpoint voltage.  Taking into account the relations for the outermost branch $\Omega_{bp} (k)=
V_{bp}/(N \sqrt{1+2\alpha(1-\cos k)})$ and $V_{bp}/N = I_{bp}/\beta$, we get
\begin{equation}
I_{bp}(\alpha, \beta, k)=\beta \sqrt{1+2\alpha(1-\cos k)}\Omega_{bp} (k,\beta).
\label{i-bp}
\end{equation}
As an example, using the expression ~(\ref{i-bp}) for $I_{bp}$,  we have plotted in Fig.\ref{3}b the
three-dimensional $\alpha\beta$-dependence of the $I_{bp}$ for two plasma modes with $k=\pi$ and $k=2\pi/5$ for
a stack with 10 IJJ. Comparing Fig.\ref{3}b with Fig.\ref{1}b, we note that the main features of the simulated
and modeled $\alpha \beta$-dependence of the $I_{bp}$  are in agreement. Using the formulas ~(\ref{i-bp}), we
have calculated the $\alpha$-dependence of the $I_{bp}$ at $\beta = 0.3$ for plasma modes with different wave
numbers $k$. The corresponding curves are presented in Fig.~\ref{3}c. We see that these results of modeling
coincide as well qualitatively with the results of simulation presented Fig.\ref{1}b. Both kinds of curves show
the same behavior.  We can see the increase in the distance between the maximums of $I_{bp}$  and their sloping
with increase in $k$ in simulated and modeled curves. Fig.~\ref{3}d shows the modeled $\beta$-dependence of
$I_{bp}$ at $\alpha = 3$, which is obtained from the resonance region data. This dependence is in agreement with
the results of simulation as well, and it demonstrates the oscillations of the $I_{bp}$, but it does not reflect
the decrease in the values of $I_{bp}$ maximums which is shown in Fig.~\ref{2}a. This is a result of the
approximations we have used to obtain the linearized equation for the Fourier component of the difference of
phase differences for  neighbor junctions.\cite{sm-sust1} The theoretical considerations which we use to model
the $\alpha\beta$- dependence of the $I_{bp}$ lead to the conclusion that there are regions on the $\alpha
\beta$-dependence of $I_{bp}$ which correspond to the creation of the LPW with a different wave number $k$ and
explain the origin of the $I_{bp}$ oscillations.
\begin{figure}[!ht]
 \centering
\includegraphics[height=60mm]{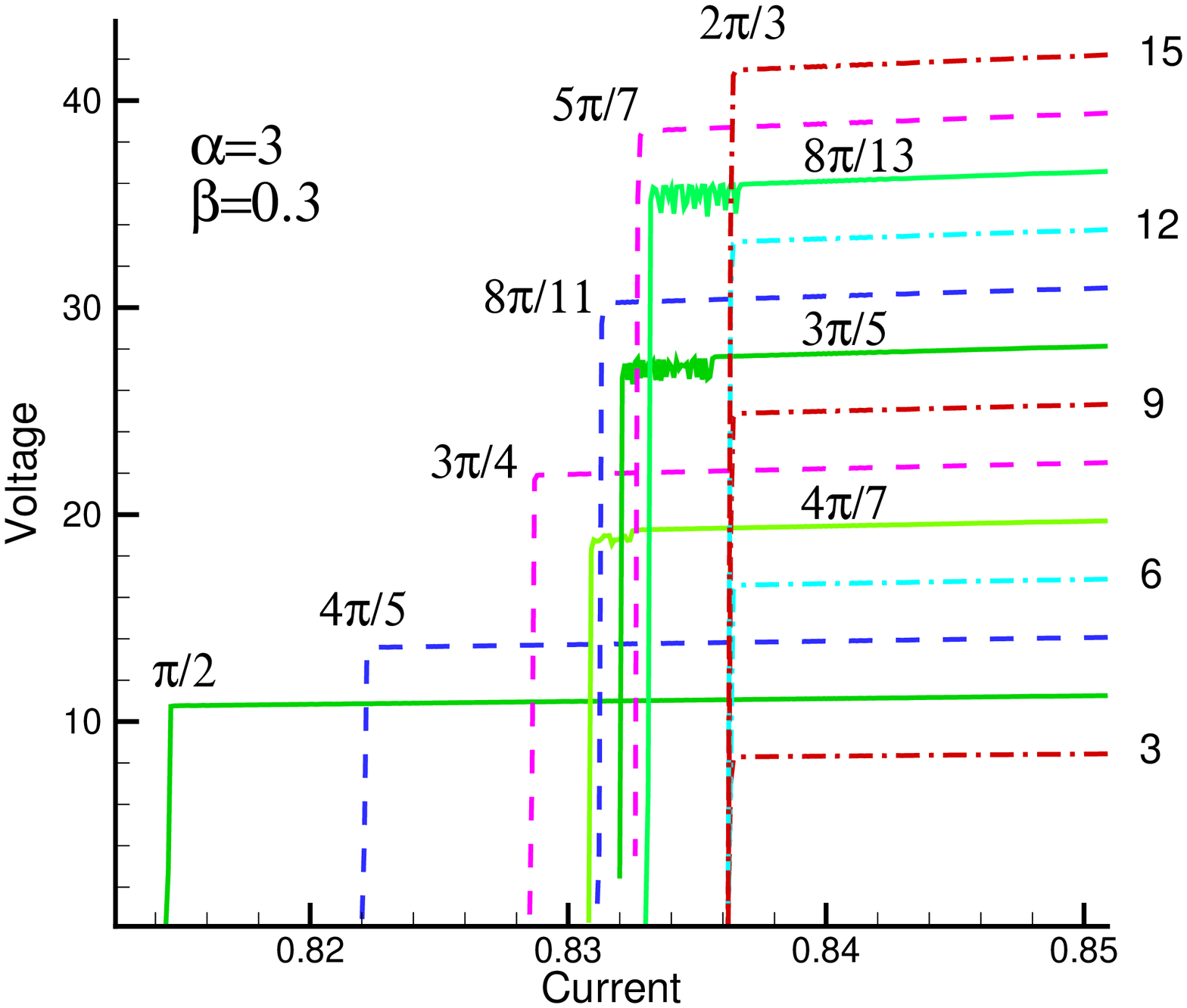}
\caption{(Color online) The simulated IVC of the outermost branch in the stacks with a different number of
junctions at $\alpha=3$, $\beta=0.3$. }
 \label{4}
\end{figure}
\begin{figure}[!ht]
 \centering
\includegraphics[height=60mm]{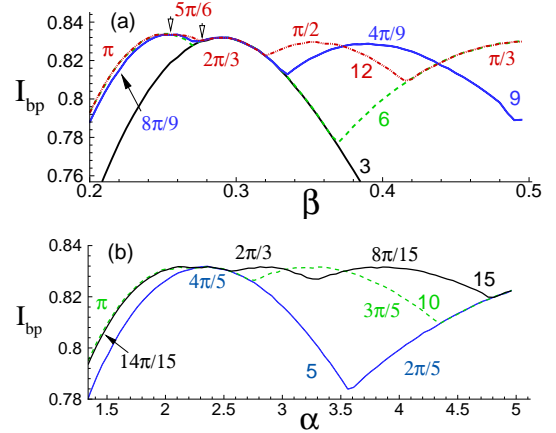}
\caption{(Color online) a) - The simulated $\beta$-dependence of the $I_{bp}$ for the stacks with 3, 6, 9 and 12
IJJ at $\alpha = 3$. The region corresponding to the creation of the  LPW mode with  wave number $k = 5\pi/6$ is
shown by arrows .  b) - The simulated $\alpha$-dependence of the $I_{bp}$ for the stacks with  5, 10 and 15 IJJ
at $\beta = 0.3$.}
 \label{5}
\end{figure}
The ideas and results presented above have strong support from the results of investigation of the $\alpha$- and
$\beta$-dependence of the $I_{bp}$ in the case of a different number of IJJ in the stack. The minimal wavelength
$\lambda$ which might be realized in the discrete lattice at periodic boundary conditions is two lattice units.
So, in the stack with N junctions,  the LPW with $k = 2\pi n/N$ may exist, where n is an integer from 1 to $N/2$
for even N and from 1 to $(N-1)/2$ for odd N. Because of the term $(1-\cos k)$ in ~(\ref{i-bp}),  the LPW with
$k$ corresponding to the highest $I_{bp}$  in the decreasing current process is created. In
Ref.\onlinecite{prb}, we showed that, at small values of $\alpha$ and $\beta$ at periodic boundary conditions
for stacks with even N, the $\pi$-mode of LPW is created, but for stacks with odd N the LPW with $k=(N-1)\pi/N$
is observed. Here we consider a case of strong coupling between junctions, and the results are different from
the previous consideration.

Fig.~\ref{4} shows the result of simulation of the outermost branch in the IVC near the breakpoint for a stack
with $\alpha=3$, $\beta=0.3$ and N from $N=3$ to $N=15$. We can see that the value of $I_{bp}$  depends on the
number N of IJJ in the stack, excluding the stack with $N = 3n$, where $n$ is an integer number. Time dependence
analysis of the charge oscillations on the S-layers shows that, at the breakpoint in the stacks with $N = 3n$,
the LPW with $k = 2\pi/3$ is created. In the stack with $N = 4$, we observe the LPW with $\lambda = 4$. We will
not touch the question concerning the breakpoint region in the IVC presented in Fig.~\ref{4}. It will be
considered in detail somewhere else. We may note another interesting  group behavior of the IVC, presented in
Fig.~\ref{4}. There is a monotonic increase of the $I_{bp}$ with $N$ for stacks with $N = 3n+1, n\geq1$. The
same monotonic behavior was observed for stacks with $N = 3n+2$. Below, we explain these results using the idea
of LPW creation at the breakpoint.

Comparison of the $\alpha$- or $\beta$-dependence of the $I_{bp}$ for stacks with a different number of IJJ give
us a simple method to determine the wave numbers $k$ of the  LPW.  Fig.~\ref{5}a shows the $\beta$-dependence of
the $I_{bp}$  at $\alpha = 3$ for the stacks with 3,6,9 and 12 IJJ. It demonstrates that, in some intervals of
$\beta$, the stacks with different N have the equal value of the $I_{bp}$. Particularly, all stacks have the
equal values of the $I_{bp}$ in some interval around $\beta = 0.3$. According to the results of modeling, for
the stack with given N, the intervals on the the curves of the $\alpha$- and $\beta$-dependence, corresponding
to the different modes of the LPW,  follow in  decreasing order in $k$. Because this interval around $\beta =
0.3$ corresponds to the region around the maximum on the $\beta$-dependence of the $I_{bp}$ for stack with $N =
3$, the second maximum for the stacks with $N = 6$ and $N = 9$, and  the third maximum for the stack with $N =
12$, we may conclude that in this interval the LPW with $k = 2\pi/3$ is created. For stacks with $N =6$ this
interval is continued until $\beta = 0.365$. Using this method of the wave number determination, which we call
$k-\alpha\beta$-method, we can determine all modes of LPW which might be created in stacks with different
parameters $\alpha$ and $\beta$ and a different number of IJJ. Particularly, we find that, on the
$\beta$-dependence, the interval (0, 0.27) and the region $\beta > 0.41$ correspond to the creation of the
$\pi$- and $ \pi/3$- modes of LPW, respectively. From the $\alpha$-dependence of the $I_{bp}$ which is presented
in Fig.~\ref{5}b for stacks with 5, 10 and 15 IJJ, we find that the interval around the maximum with $2.35$ and
the region $\alpha > 4.82$ correspond to the creation of the  $4\pi/5$- and $ \pi/5$- modes of LPW,
respectively.

Using the $k-\alpha\beta$-method, we find the values of $k$ for IVC presented in Fig.~\ref{4}. In the stacks
with $N = 3n$ (dash-dotted curves in Fig.~\ref{4}), the LPW with the same wave number $k = 2\pi/3$ are created.
For the stacks with $N = 3n+1$ (solid curves), we obtain  $k = 2(N-1)\pi/3N$. This value limits to $2\pi/3$ with
an increase in $N$ from the side of smaller values of $k$. In the stacks with $N = 3n+2$ (dash curves), we get
$k = 2(N+1)\pi/3N$, which limits to $2\pi/3$ from the side of bigger values of $k$. So the idea of the LPW
creation at the breakpoint explains the group behavior of IVC in Fig.~\ref{4}. The value of $I_{bp}$ depends on
$k$ but does not depend on $N$ at chosen parameters $\alpha$ and $\beta$; i.e., the creation of the same mode in
the stacks with different $N$ leads to the same value of $I_{bp}$. So we may predict a different
commensurability manifestation in the IVC of stacks with a different number of IJJ. This is a generalization of
the commensurability effect we have observed in Ref.\onlinecite{prb} at small $\alpha$ and $\beta$.

As summary, we showed that coupling between junctions changes crucially the dependence of the return current on
a dissipation parameter. Particularly, it leads to the appearance of the plateau on the $\beta$-dependence of
the BPC on the outermost branch and the oscillation of the BPC  as a function of $\beta$. Using the idea that at
the breakpoint the parametric resonance is approached and a longitudinal plasma wave is created, we modeled the
$\alpha$- and $\beta$-dependence of the BPC and obtained good agreement with the results of the numerical
simulation. We demonstrated that the study of the $\alpha$- and $\beta$-dependence of the BPC for the stacks
with a different number of IJJ gives us the instrument to determine the wave number of the LPW.

We thank  N.M.Plakida, Y.Sobouti, M.R.H.Khajehpour for support of this work.


\section*{References}
\begin{thebibliography}{10}

\bibitem{muller} R. Kleiner, F. Steimmeyer, G. Kunkel and P. Muller, Phys. Rev. Lett. {\bf68}, 2394 (1992); G. Oya, N. Aoyama, A. Irie, S. Kishida, and H. Tokutaka, Jpn. J. Appl. Phys., {\bf31}, L829 (1992).
\bibitem{schmidt}D. E. McCumber, J.Appl.Phys. 39, 3113 (1968); W. C. Steward, Appl.Phys.Lett. {\bf12}, 277 (1968).
\bibitem{sm-sust1}Yu. M. Shukrinov, F.Mahfouzi, Supercond. Sci.Technol., {\bf19}, S38-S42 (2007).
\bibitem{prb}Yu. M. Shukrinov, F.Mahfouzi, N. F. Pedersen, Phys. Rev. B {\bf 75}, 104508 (2007).
\bibitem{machida04} M. Machida, T. Koyama, Phys. Rev. B {\bf 70}, 024523  (2004).
\bibitem{machida00} M. Machida, T. Koyama, A. Tanaka and M. Tachiki, Physica {\bf C330}, 85 (2000)
\bibitem{sm-physC2}Yu. M. Shukrinov, F. Mahfouzi, P. Seidel.  Physica {\bf C449}, 62 (2006).
\bibitem{koyama96} T. Koyama and M. Tachiki, Phys. Rev. B {\bf 54},  16183  (1996)
\bibitem{matsumoto99}H. Matsumoto, S. Sakamoto, F. Wajima, T. Koyama, M. Machida, Phys. Rev. B {\bf 60}, 3666 (1999)
\bibitem{sm-physC1}Yu. M. Shukrinov and F. Mahfouzi,  Physica  {\bf C434}, 6 (2006).


\end{thebibliography}
\end{document}